\documentclass{aa}  

\usepackage{graphicx}
\usepackage{txfonts}
\usepackage{lipsum}
\usepackage{subcaption}         % necessary for continued figures, example in section 3
\usepackage{lscape}             % to rotate a single page table, example in appendix.
\usepackage{placeins}           % useful with \FloatBarrier, to keep 
\usepackage{hyperref}
\usepackage{color}

\newcommand{\IGNORE}[1]{}

\newcommand{\bea}{\begin{eqnarray}}
\newcommand{\eea}{\end{eqnarray}}

\begin{document}

\title{Planar Three-Body Problem: \\ theoretical predictions and simulation results}

   \subtitle{}

   \author{Yogesh Dandekar\inst{1}\corrauth{yogesh.dandekar@biu.ac.il}        % use \corrauth for the corresponding author
        \and Ethan Springer\inst{1}\email{}
        \and Shoval Zard\inst{1}\email{}
        \and Alessandro Alberto Trani\inst{2,3}\email{}
        \and Barak Kol\inst{4}\email{}
        \and Ofek Birnholtz\inst{1}\email{}
        }

   \institute{Department of Physics, Bar-Ilan University, Ramat Gan 5290002, Israel \and Department of Astronomy, University of Concepción, Avenida Esteban Iturra s/n Casilla 160-C Concepción, Chile \and INFN, Sezione di Trieste, I-34127, Trieste, Italy \and Racah Institute of Physics, Hebrew University, Jerusalem 91904, Israel}

   \date{Received M DD, YYYY}
 
  \abstract
   {}
   {We investigate the statistical properties of the non-hierarchical planar three-body problem, relevant to astrophysical systems with nearly planar dynamics, including protoplanetary disks and AGN disks.} 
   {Using the Flux-based statistical theory of the three-body problem, previously studied in three dimensions, we derive theoretical predictions for several statistical observables in the planar case. Focusing on ergodic disintegration outcomes, we compare these predictions directly with numerical simulations and contrast the results with the corresponding unconstrained three-dimensional case. Our analysis is based on one million simulations for each of eight distinct mass sets.}
   {We examine escape probabilities, properties of marginal escape events, eccentricity distributions, lifetime distributions, and relative prevalence of prograde and retrograde escapes. While several of the theoretical predictions are in good agreement with the simulations, others exhibit unexpected discrepancies. We discuss possible explanations for these deviations.}
   {}


   \maketitle
%%%%%%%%%%%%%%%%%%%%%%%%%%%%%%%%%%%%%%%%%%%%%%%%%%%%%%%%%%%%%%
\nolinenumbers
\section{Introduction}  

The three-body problem is one of the fundamental challenges in physics, with applications spanning a wide range of astrophysical systems. A general three-body problem considers the motion of three bodies under their mutual gravitational interactions. This paper focuses on a particular case of the three-body problem in which three bodies with comparable (non-hierarchical) masses evolve under Newtonian gravity, with the total energy of the system being negative. The three-body problem exhibits chaotic behavior, i.e., small perturbations in the initial conditions diverge exponentially with time. Consequently, for generic initial conditions, the three-body system eventually decays into a binary and an escaping body.

Due to the chaotic nature of the three-body problem, attempts to obtain analytic solutions for the trajectories and outcomes are highly limited. Nevertheless, a statistical approach has proven suitable for analyzing the outcomes. In this framework, probability distributions of various statistical quantities of interest are considered. These statistical properties depend only on the values of the three masses, the total energy, and the total angular momentum of the system. In this paper, we primarily focus on the Flux-based statistical theory introduced in \citep{2021CeMDA.133...17K} (see also \citep{2023CeMDA.135...29K}).
For developments of other statistical theories see \citep{1976MNRAS.176...63M,1976MNRAS.177..583M,1978MNRAS.184..119N,2006tbp..book.....V,2019Natur.576..406S,2021PhRvX..11c1020G,2022MNRAS.517.3838L,2023ApJ...952..103Z,2024MNRAS.531..739G,2024MNRAS.533..486G,2024MNRAS.535L..26K,2023ARep...67..742K}.

\begin{figure}[t]
	\centering
	\includegraphics[width=\hsize]{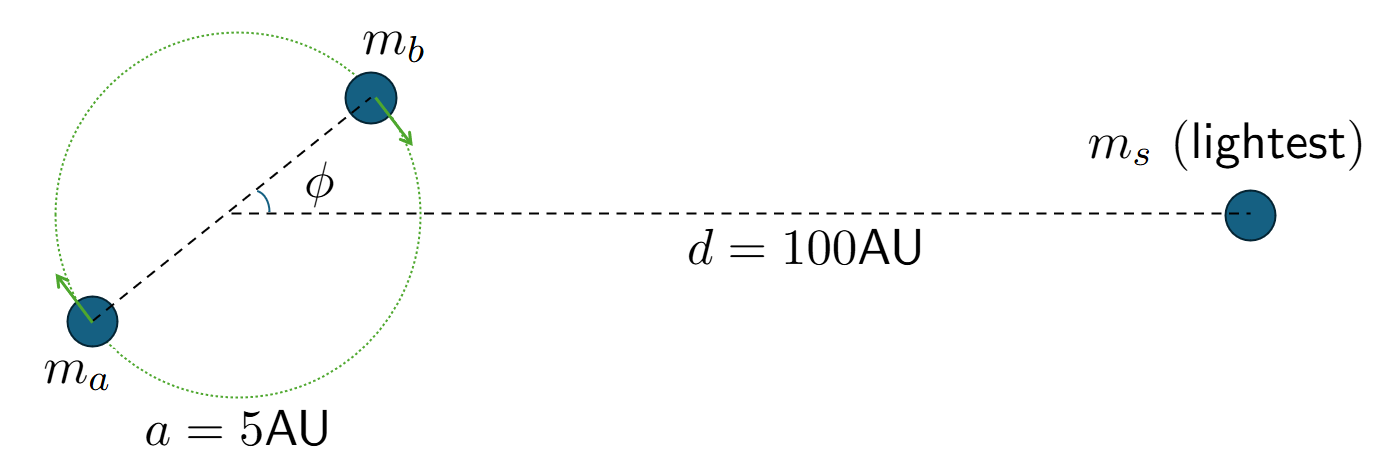}
	\caption{Initial conditions setup for the planar three-body problem. Masses are denoted by $m_i$, semi-major axis of binary is denoted by $a$, and the distance from the binary center of mass to the single body is denoted by $d$. Binary center of mass and the single body are initially at rest. $\phi$ is chosen randomly from a uniform distribution over the interval $[0,2\pi]$.}
	\label{inco}
\end{figure}

 The three-body problem can be formulated in either three or two dimensions\footnote{In the two-dimensional case, the initial velocities of the three bodies are constrained to the plane defined by their initial positions. Without spins, the subsequent evolution preserves the plane of motion.}. In this paper, we analyze the planar three-body problem using both current theoretical understanding and numerical computations (our initial conditions setup appears in Figure \ref{inco}). The planar problem is intrinsically interesting both in its own right and for comparison with the corresponding three-dimensional results,
providing more test cases for the general statistical theory.

\begin{figure*}[!t]
	\centering
    \sidecaption
	\includegraphics[width=0.7\linewidth]{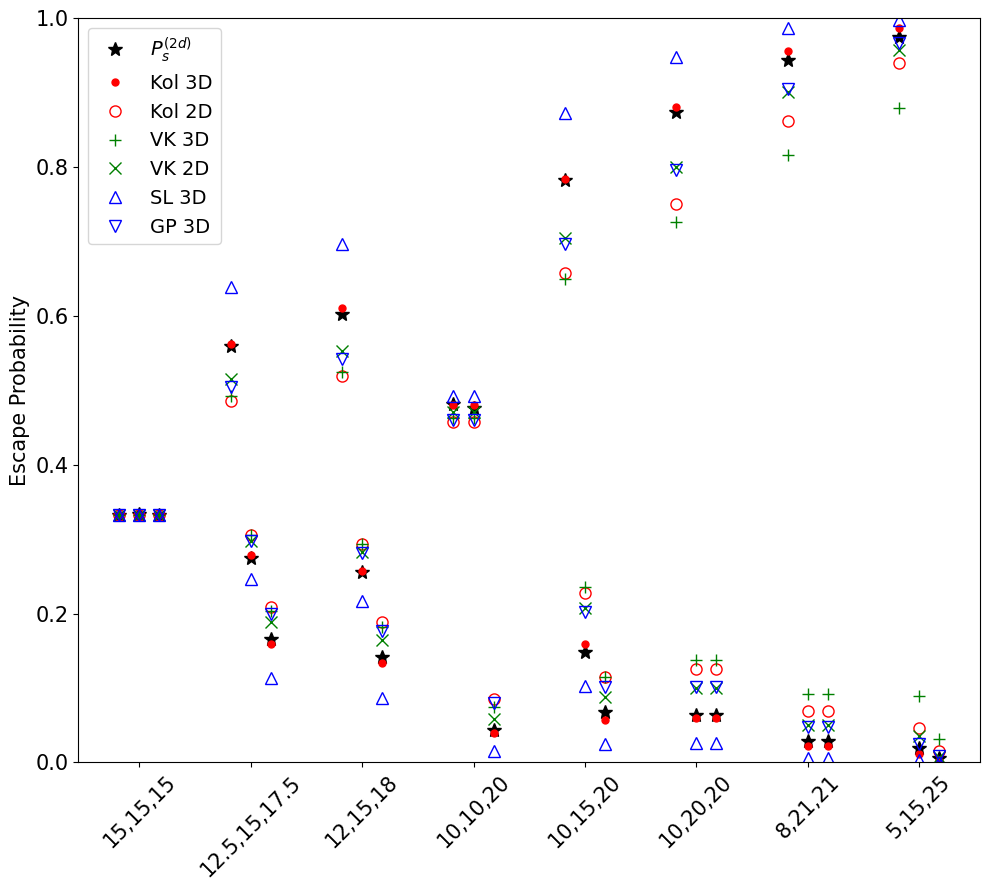}
	\caption{Escape probabilities for all mass sets. For each mass set, the escape probabilities are ordered according to the lightest, intermediate, and heaviest body from left to right. $P_{s}^{(2d)}$ denotes the escape probabilities measured from the planar (2D) simulations after applying the ergodic cut. Kol 2D (Eq. \eqref{ep2}) and Kol 3D (Eq. \eqref{ep3}) denote the corresponding predictions of the Flux-based theory \citep{2021CeMDA.133...17K}. VK 2D and VK 3D denotes the predictions derived in \citep{2006tbp..book.....V}. SL 3D and GP 3D denotes the predictions derived in \citep{2019Natur.576..406S}, \citep{2021PhRvX..11c1020G}, respectively. Surprisingly, the planar simulation results $P_{s}^{(2d)}$ agree with Kol 3D predictions (under emissivity-blind approximation) at the $1\%$ level, and deviate substantially from all the other predictions.\\
    \vspace{40pt}
    }
	\label{epfig}
\end{figure*}

In addition, the present study is particularly relevant in at least two important astrophysical contexts:
the first is the formation of planets in protoplanetary disks (see, e.g., \citep{Armitage2017, 2022ASSL..466....3R}). Protoplanetary disks are typically thin, making the mutual interactions between protoplanets approximately planar. Although planet formation is inherently an N-body problem, understanding the statistical properties of the planar three-body problem represents the first nontrivial step toward developing statistical descriptions of planar N-body interactions.\footnote{See \citep{2016MNRAS.463.3311L, 2024MNRAS.528..198B} for the work on four-body interactions in three dimensions, as well as Y. Zak, H. Perets, in prep.} The present work therefore provides a foundation for extending statistical approaches to more realistic models of planet formation and dynamical evolution in protoplanetary disks.

A second application arises in the dynamics of black-hole triples in AGN disk environments. Recent studies have shown that binary--single interactions in AGN disks provide a mechanism for the formation and merger of binary black holes (see, e.g., \citep{2020ApJ...898...25T}). These nearly coplanar interactions lead to qualitative differences in the properties of merging binary black holes compared to the fully three-dimensional interactions characteristic of stellar clusters (see, e.g., \citep{2022Natur.603..237S}). In this context, the present work can be viewed as an initial step toward developing statistical descriptions of three-body interactions involving black holes in AGN disk environments.

To test the statistical theories of the three-body problem, a large number of numerical simulations are required. Such simulations enable the measurement of outcome probability distributions, which can then be compared with theoretical predictions from statistical theories of the three-body problem. See \citep{2020MNRAS.497.3694M,2021MNRAS.506..692M,2024CeMDA.136....4M,2024A&A...689A..24T} for various statistical results obtained through such simulations.

In \citep{2021MNRAS.506..692M}, hereafter referred to as MKTL21, the authors performed one million three-body simulations for each of the eight chosen mass sets in the three-dimensional case. They measured various statistical quantities from these simulations and compared the results with theoretical predictions.

In this paper, we derive theoretical predictions for the planar problem using the Flux-based statistical theory of \citep{2021CeMDA.133...17K}. We also perform one million simulations for the same set of eight mass sets chosen in MKTL21, allowing for direct comparison with the three-dimensional simulation results. The theoretical predictions are then compared with the simulation outcomes. In figure \ref{epfig}, we present one of the key results of this paper, which compares the simulation results with various theoretical predictions of the escape probabilities for the eight mass sets. Such results will be discussed in detail in Section \ref{secres}.

This paper is organized as follows. In Section \ref{secthe}, we derive theoretical predictions for various statistical quantities within the framework of the flux-based theory for the planar problem. In Section \ref{secsim}, we describe the simulation setup, including the initial conditions, a brief overview of the numerical code, and the data recorded for each simulation. In Section \ref{secres}, we present the simulation results and compare them with theoretical predictions. Finally, in Section \ref{secsum}, we summarize the main findings and discuss possible directions for future work.

\section{Theoretical Predictions}\label{secthe}

In this section, we derive various predictions of the Flux-based theory introduced in \citep{2021CeMDA.133...17K}. We begin with an overview of the central formalism of the Flux-based theory that is directly relevant to this work. We then derive the total phase-space volume flux for the planar problem. Using the Flux-based theory, we derive predictions for the escape probabilities, the distributions of outer angular momentum for marginal escapes, the eccentricity distributions of the inner and outer binaries, and the relative prevalence of prograde and retrograde escapes. Finally, we discuss theoretical expectations for the lifetime distribution. Appendix \ref{symsec} summarizes the symbols frequently used throughout this paper.

\subsection{Three-body problem}
The Hamiltonian of the three-body problem is given by
\begin{equation}
    H := \sum_{i=1}^{3} \frac{p_i^2}{2m_i} - \sum_{i<j} \frac{G m_i m_j}{r_{ij}} 
\end{equation}
where $m_i$ and $p_i$ denote the three masses and their corresponding momenta. $r_{ij}$ denotes the distance between masses $m_i$ and $m_j$. The conserved quantities are the total energy, total angular momentum, and total linear momentum, denoted by $E$, $\vec{L}$, and $\vec{P}$, respectively.

A non-hierarchical three-body interaction with generic initial conditions and negative total energy eventually disintegrates into a bound binary and an escaping body. Visualizations of the simulations reveal that the three-body systems exhibit both regular and chaotic motion. 

The regular motion consists of incoming and outgoing asymptotic motion, as well as sub-escape excursions. Sub-escape excursions are defined as motions in which a bound binary and the single body begin moving away from each other, attain a large separation, but eventually fall back toward each other due to insufficient kinetic energy. These excursions can have arbitrarily long durations and arbitrarily large binary-single separations. 

The chaotic motion consists of a sequence of scramble states and leads the system toward losing memory of the initial conditions preceding the chaotic phase. In a scramble state, the three bodies are approximately equidistant from one another and are in approximate equipartition of energy. Therefore, these states are non-hierarchical. %A typical scramble state lasts approximately $2$--$3$ years, as observed from simulations.

Apart from the incoming and outgoing asymptotic motion, a typical three-body interaction consists of a sequence of chaotic motion and sub-escape excursions combined in a random manner with varying proportions.  

When the binary and the single body are sufficiently well separated, the three-body system can be decomposed into an {\it inner binary} and an {\it outer binary}. The inner binary is defined as the bound binary system. The outer binary system is the effective hierarchical system composed of the single body and a fictitious body located at the center of mass of the inner binary, with mass equal to the total mass of the inner binary. The outer binary describes the relative motion between the bound binary and the single body.

\subsection{Flux-based theory}\label{thint}

Flux-based theory is a statistical framework for the three-body problem developed in \citep{2021CeMDA.133...17K}. It is applicable to non-hierarchical systems with negative total energy. In this theory, the {\it flux} of phase-space volume governs the statistical predictions of observable outcomes. In contrast to previous statistical approaches that associate probabilities with phase-space volume, Flux-based theory associates probabilities with the {\it flux} of phase-space volume. An intuitive analogy is that of a leaky container representing an open chaotic system, such as the three-body problem. In this analogy, the flux of the phase-space volume is represented by air escaping from the leaky container, and the regularized phase-space volume is represented by the container's volume.

A central prediction of the Flux-based theory is
\begin{equation}\label{fbte}
	d\Gamma (u) = \frac{{\cal E}(u) dF(u)}{\bar\sigma(E,L)}
\end{equation}
 where $u$ denotes the asymptotic state parameters,
 which in the planar case consist 
 % consisting 
  of the total energy, angular momentum, and pericenter angle for the inner binary ($E_B$, $L_B$, $\phi_B$), together with the corresponding quantities for the outer binary ($E_F$, $L_F$, $\phi_F$). 
 
 $d\Gamma$ is the differential decay rate out of the ergodic region. It is defined as the probability per unit time that a system undergoing ergodic motion decays into a regular state with asymptotic state parameters $u$. $dF$ denotes the differential flux of phase-space volume into asymptotic states. The expression for $dF$ was derived in closed form in \citep{2021CeMDA.133...17K}. $\bar{\sigma}$ is the regularized phase-space volume of the three-body problem and was derived in \citep{2022CeMDA.134...55D,2025CeMDA.137...32D}.
 
 Finally, ${\cal E}$ denotes the emissivity, defined as the probability that an asymptotic state with parameters $u$ originated from a chaotic state. It characterizes the partition of phase space into regular and chaotic regions. Since ${\cal E}$ is not known analytically, it must be measured using three-body simulations, as done in \citep{2024CeMDA.136....4M}. To circumvent this limitation, we restrict our attention to statistical quantities that are only weakly dependent on ${\cal E}$ or make physically motivated assumptions about ${\cal E}$.

\subsection{Derivation of Flux}\label{dfs}

Since the derivation presented in this section closely follows the corresponding derivation for the three-dimensional case, we refer the reader to Section 3.2 of \citep{2021CeMDA.133...17K}, where the 3D derivation is presented.

The microcanonical phase-space volume for the central force problem in the planar situation is given by
\begin{equation}
    d\sigma^{(2d)}(E,L) = \int d^2p~ d^2r~ \delta(H-E) \delta(J-L)
\end{equation}
where $H$ and $J$ denote the expressions for Hamiltonian and Total Angular momentum, respectively. $E$ and $L$ denote the values of the Total Energy and Total Angular momentum, respectively. 

Following manipulations analogous to those used in the 3D case (see Eq. (3.17) of \citep{2021CeMDA.133...17K}), we obtain
\begin{equation}\label{pst}
    \begin{split}
        d\sigma^{(2d)}(E,L) &= \int dr~ d\phi~ dp_r~ dJ~ \delta(H-E) \delta(J-L) \\
        &= d\phi~ \int dr~ dp_r~ \delta(H_{\text{eff}}-E) \\
        &= d\phi~ dt_r
    \end{split}
\end{equation}
where polar coordinates have been used. In the first line, Eq. (3.16) of \citep{2021CeMDA.133...17K} has been used. In the second line, integration over $J$ has been performed and the effective Hamiltonian has been defined as $H_{\text{eff}} = p_r^2/(2\mu) + L^2/(2\mu r^2) + V(r)$. In the third line, $dt_r$ denotes the time element of the radial problem. 

For bound motion, integrating Eq. \eqref{pst} over $t_r$ yields
\begin{equation}\label{bms}
    d\sigma_B^{(2d)}(E,L) = d\phi~ T(E)
\end{equation}
where $T(E)$ denotes the binary period. 

Using Eq. \eqref{pst}, the phase-space volume flux is given by 
\begin{equation}\label{fmf}
    dF_F^{(2d)}(E,L) = \lim_{R\rightarrow \infty} d\sigma^{(2d)}~ \delta(r-R) \dot r  = d\phi
\end{equation}
Similar to Eq. (3.21) of \citep{2021CeMDA.133...17K}, Eqs. \eqref{bms} and \eqref{fmf} can be rewritten in a more convenient form as
\begin{equation}\label{rep}
    \begin{split}
        d \tilde \sigma_B^{(2d)} (E,L) &= d \sigma_B^{(2d)} (E,L)~ dE~ dL \\
        d \tilde F^{(2d)}_F (E,L) &= d F^{(2d)}_F (E,L)~ dE~ dL
    \end{split}
\end{equation}
Now consider the planar three-body problem. As discussed in Section \ref{thint}, the system can be decomposed into an inner binary and an outer binary when the binary and the single body are sufficiently well separated. 

For the unbound motion of the outer binary, using Eqs.\eqref{fmf} and \eqref{rep}, we obtain
\begin{equation}\label{flt}
    d \tilde F_F^{(2d)} (E_F,L_F) = d\phi_F~ dE_F~ dL_F
\end{equation}
For the bound Kepler problem of the inner binary, using Eqs. \eqref{bms} and \eqref{rep}, we obtain
\begin{equation}\label{psb}
    \begin{split}
        d \tilde \sigma_B^{(2d)}(E_B,L_B) &= d\phi_B~ dE_B~ dL_B~ T(E) \\
        &= d\phi_B~ dE_B~ dL_B \left( 2\pi \frac{\sqrt{k_B}}{(-2E_B)^{3/2}} \right)
    \end{split}
\end{equation}
where $k_B=\mu_B\alpha_B^2$, with $\mu_B:=m_am_b/(m_a+m_b)$ denoting the reduced mass and $\alpha_B:=Gm_am_b$ denoting the potential strength constant of the binary.

Note that both $L_B$ and $L_F$ can take positive or negative values. The total phase-space volume flux for the three-body problem can be written using the decomposition into inner and outer binaries as
\begin{equation}\label{f2d}
    \begin{split}
        F^{(2d)} &= \int d \tilde F_F^{(2d)}~ d \tilde \sigma_B^{(2d)}~ \delta(E_B+E_F-E)~ \delta(L_B+L_F-L) \\
        &= \int d\phi_F~ dE_F~ dL_F~ d\phi_B~ dE_B~ dL_B \left( 2\pi \frac{\sqrt{k_B}}{(-2E_B)^{3/2}}  \right) \times \\ &\quad\quad\quad   \times \delta(E_B+E_F-E)~ \delta(L_B+L_F-L) \\
        &= (2\pi)^3 \int dL_s~ dL_B \\
        &= (2\pi)^3 \int (L_0 - |L_B|) dL_B \\
        &= (2\pi)^3 L_0^2
    \end{split}   
\end{equation}
In the second line, Eqs. \eqref{flt} and \eqref{psb} have been substituted. In the third line, integration has been performed over $E_F$, $L_F$, $\phi_B$, and $\phi_F$, and the notation 
\begin{equation*}
    dL_s = \frac{\sqrt{k_B}}{(-2E_B)^{3/2}}dE_B
\end{equation*}
has been introduced, where $L_s\equiv\sqrt{k_B/(-2E_B)}$ denotes the maximum possible value of $|L_B|$ for a given value of $E_B$. 

In the fourth line, integration over $L_s$ has been performed, and 
\begin{equation}
    L_0 \equiv \sqrt{k_B/(-2E)}
\end{equation}
has been introduced as the maximum possible value of $|L_B|$ given $E_B=E$. Finally, the integration over $L_B$ is carried out from $-L_0$ to $+L_0$. 

Therefore, the total phase-space volume flux for the planar three-body problem is
\begin{equation}\label{fpl}
        F^{(2d)} = (2\pi)^3 L_0^2
\end{equation}
Note that the result in Eq. \eqref{fpl} differs from the corresponding expression for the 3D case derived in Section 4.2 of \citep{2021CeMDA.133...17K},
\begin{equation}\label{f3d}
    F^{(3d)} =  
    \begin{cases}
   \frac{(2\pi)^4}{3} L_0^2 \left(3-3\frac{L}{L_0}+\frac{L^2}{L_0^2}\right) & \text{if } 0\leq L\leq L_0 \\
    \frac{(2\pi)^4}{3} \frac{L_0^3}{L}  & \text{if } L_0 \leq L
\end{cases}
\end{equation}
The dimensions of Eq. \eqref{fpl} are identical to those of Eq. \eqref{f3d}, as expected:
\begin{equation}
    \left[F^{(3d)}\right] = \left[F^{(2d)}\right]
\end{equation}
Furthermore, in the limit $L\rightarrow0$, the following relation between Eqs. \eqref{fpl} and \eqref{f3d} is satisfied, as expected:
\begin{equation}
    \lim_{L\rightarrow 0}~ F^{(3d)} = (2\pi)~\lim_{L\rightarrow 0}~  F^{(2d)}
\end{equation}

\subsection{Escape probabilities}\label{pes}

The flux-based theory predicts the escape probabilities of the three bodies through the relative values of their corresponding total decay rates. The total decay rates are obtained from Eq. \eqref{fbte} by integrating over the outcome parameters $u$. We assume that emissivity does not significantly affect the relative values of the total decay rates, since its effects are averaged over. Under this emissivity-blind approximation and using Eq. \eqref{fpl}, we obtain
\begin{equation}\label{ep2}
	P^{(2d)}_s \propto L_0^2 
\end{equation}
where $P_s^{(2d)}$ denotes the escape probability of body $s$ in the planar system, and the value for $L_0$ is chosen appropriately for the escaper.  

Now consider the 3D case with $L\geq L_0$. We restrict to $L\geq L_0$ because of the manner in which our simulations are set up. In this case, the corresponding escape probability for body $s$ under the emissivity-blind approximation is given by (see section 4.2 of \citep{2021CeMDA.133...17K})
\begin{equation}\label{ep3}
    P^{(3d)}_s \propto L_0^3 
\end{equation} 

The predictions Eq. \eqref{ep2} and Eq. \eqref{ep3} are compared with simulation results in Section \ref{epds}.

\subsection{Marginal escapes}\label{mes}

We consider the subset of three-body systems in which the final escape is marginal, i.e., the outer binary possesses only a small positive total energy $E_F$. 

Changing variables from $L_B$ to $L_F$ in Eq. \eqref{f2d}, we obtain
\begin{equation}
\begin{split}
    F^{(2d)} &= (2\pi)^3 \int dL_s~ dL_F \\
    &= (2\pi)^3 \int (L_0-|L-L_F|)~ dL_F
\end{split}
\end{equation}
 Consider the case $L\geq L_0$, which is relevant for our simulations. For escapes occurring by a narrow margin, $L_F$ is only slightly larger than the threshold value $L_{F,c}\equiv L-L_0$. In this regime, the distribution of $F^{(2d)}$ over $L_F$ becomes
\begin{equation}
    dF^{(2d)} \propto (L_F-L_{F,c})~ dL_F
\end{equation}
The emissivity is expected to be large and smoothly varying when the escaper is only marginally unbound from the binary. Consequently, the $L_F$ distribution for marginal escapes is expected to be insensitive to the emissivity factor. Using Eq. \eqref{fbte}, we obtain
\begin{equation}\label{lf2d}
    dP_s^{(2d)} \propto (L_F-L_{F,c})~ dL_F
\end{equation}
while in the 3D case for $L\geq L_0$ we have (see section 4.1 of \citep{2021CeMDA.133...17K})
\begin{equation}\label{lf3d}
    dP_s^{(3d)} \propto (L_F-L_{F,c})^2~ dL_F
\end{equation}
The prediction in Eq. \eqref{lf2d} is compared with simulation results in Section \ref{meds}.

\subsection{Eccentricities}\label{eds}

We consider the distributions of the eccentricity of the remnant inner binary ($0\leq e_B\leq1$) and the eccentricity of the remnant outer binary ($e_F\geq1$).

For the inner binary, using Eq. \eqref{f2d} and changing variables to $e_B^2\equiv1+\frac{2E_BL_B^2}{k_B}$, we obtain
\begin{equation}
	dF^{(2d)} \propto \frac{e_B}{\sqrt{1-e_B^2}}~de_B
\end{equation}
while in the 3D case for $L\geq L_0$ we have (see Section 4.2 of \citep{2021CeMDA.133...17K})
\begin{equation}\label{ibe3}
	dF^{(3d)} \propto e_B~de_B
\end{equation}

For the outer binary, using Eq. \eqref{f2d} and changing variables to $e_F^2 \equiv1+\frac{2E_FL_F^2}{k_F}$, we obtain
\begin{equation}
	dF^{(2d)} \propto \frac{e_F}{\sqrt{e_F^2-1}}\left.\sqrt{L_0^2-L_\epsilon^2}\right|_{L_\epsilon=L_{\epsilon,\text{max}}}^{L_\epsilon=L_{\epsilon,\text{min}}}~de_F
\end{equation}
where $L_{\epsilon,\text{max}}$ and $L_{\epsilon,\text{min}}$ are solutions to the following quartic equations
\begin{equation}\label{limeqs}
	\begin{split}
		\left(L-L_{\epsilon,{\text min}}\right)^2\left(\frac{1}{L_{\epsilon,{\text min}}^2}-\frac{1}{L_0^2}\right) &= \left(e_F^2-1\right)\frac{k_F}{k_B} \\			
        \left(L+L_{\epsilon,{\text max}}\right)^2\left(\frac{1}{L_{\epsilon,{\text max}}^2}-\frac{1}{L_0^2}\right) &= \left(e_F^2-1\right)\frac{k_F}{k_B} 
	\end{split}
\end{equation}

Assuming the validity of the emissivity-blind approximation, the theoretical predictions for the eccentricity distributions of the remnant inner and outer binaries for the planar case are given by
\begin{equation}\label{ibe}
dP_s^{(2d)} \propto \frac{e_B}{\sqrt{1-e_B^2}}~de_B
\end{equation}
and
\begin{equation}\label{obe}
dP_s^{(2d)} \propto \frac{e_F}{\sqrt{e_F^2-1}}\left.\sqrt{L_0^2-L_\epsilon^2}\right|_{L_\epsilon=L_{\epsilon,\text{max}}}^{L_\epsilon=L_{\epsilon,\text{min}}}~de_F
\end{equation}
respectively. \footnote{Note that there is useful choice of variables, $e_B = \sin \theta$, which transforms the $e_B$ distribution as $dP_s^{(2d)} \propto \sin\theta ~d\theta$.}
% For $e_F$ close to unity, $dP_s^{(2d)}\propto de_F$, and 
 For large $e_F$, $dP_s^{(2d)}\propto e_F^{-3}~de_F$, plus higher order terms.

The predictions given by Eq. \eqref{ibe} and \eqref{obe} are compared with the simulation results in Section \ref{ioes}.

\subsection{Lifetime distributions}\label{lds}

Since sub-escape excursions can last arbitrarily long, the late-time region of the lifetime distribution is expected to be dominated by long sub-escape excursions rather than ergodic motion. Using Kepler's third law, it can be argued (see, e.g., \citep{1983AAO_orig,1983AAO_trans,1993ApJ...403..256H,2021MNRAS.506..692M}) that the lifetime distribution follows a power-law decay in the late-time region. Specifically, the differential lifetime distribution is expected to scale as $\tau_D^{-5/3}$, where $\tau_D$ denotes the disintegration time.

Escapes originating from the ergodic region give rise to an exponentially decaying lifetime distribution, characteristic of memoryless systems. The systems with shorter lifetimes are expected to contain a mixture of ergodic motion and sub-escape excursions. Consequently, the early-time lifetime distribution is expected to consist of both exponential and power-law components. These theoretical expectations are tested against numerical simulations in Sections \ref{ldds} and \ref{sefs}.

\subsection{Prograde and Retrograde escapes}\label{prs}

For planar three-body systems, we classify the disintegrations into two subsets according to whether the angular momentum of the remnant inner binary ($L_B$) is prograde or retrograde relative to the direction of the total angular momentum ($L$).

It can be shown from Eq. \eqref{f2d} that the theory does not distinguish between prograde and retrograde escapes, under the emissivity-blind approximation. In particular, the probability of a prograde escape is predicted to be equal to that of a retrograde escape. Consequently, the escape probabilities of the individual bodies are expected to remain unchanged under the prograde and retrograde classifications. These expectations are compared with simulation results in Section \ref{prss}.

\section{Simulation Setup}\label{secsim}

In this section, we describe the simulation setup in detail. We first discuss the choice of initial conditions, followed by details of the numerical code, including a discussion of numerical errors. Finally, we summarize the quantities recorded from each simulation run. The simulations are set up using a procedure closely following that of MKTL21 \citep{2021MNRAS.506..692M}. 

\subsection{Initial conditions}

We adopt units such that $G=4\pi^2$, mass is measured in solar masses ($M_\odot$), and distance is measured in astronomical units (AU), implying that time is measured in years. Note that MKTL21 instead adopts $G=1$, $M_\odot$, and AU.

We work in the center-of-mass frame and consider an initial configuration consisting of a bound binary ($m_a,m_b$) in a circular orbit and a distant single body ($m_s$), with both the single body and the binary center of mass initially at rest. The lightest body among the three is chosen to be the single body ($m_s$). The initial semi-major axis of the binary ($a$) is set to 5 AU, while the initial separation between the binary center of mass and the single body ($d$) is set to 100 AU. The binary and the single body subsequently fall toward one another under Newtonian gravity.

For the planar case, we perform one million simulations by selecting the initial binary phase randomly from a uniform distribution over the interval $[0,2\pi]$. The inclination of the single body with respect to the binary plane is fixed to be 0. Figure \ref{inco} visualizes the choice of initial conditions. For the 3D case, we perform one million simulations by selecting the initial binary phase randomly from a uniform distribution over $[0,2\pi]$ and the inclination of the single body with respect to the binary plane randomly from a uniform distribution over $[0,\pi]$.

This choice of initial conditions ensures that a large fraction of the simulated systems undergo chaotic evolution, which is precisely the regime in which statistical theories are expected to be applicable.

These initial conditions imply the following expressions for the total energy\footnote{The expression for $E$ in Eq. \eqref{ine} is an approximation. However, the exact values of $E$ differ only slightly, making this approximation sufficiently accurate for our analysis.} and the total angular momentum
\begin{equation}\label{ine}
	E = -\frac{G m_a m_b}{2a} - \frac{G m_s (m_a+m_b)}{d} 
\end{equation}
\begin{equation}\label{inl}
	L = \sqrt{G}m_am_b\sqrt{\frac{a}{m_a+m_b}}
\end{equation}
Note that for these initial conditions it can be verified that $L\geq L_0$, where $L_0\equiv\sqrt{k_B/(-2E)}$ represents the maximum angular momentum that a binary can possess while remaining circular and having total energy $E$. Note that there are three possible values of $L_0$ corresponding to the three possible binary combinations. 

We consider the eight mass sets used in MKTL21, enabling a direct comparison of the planar results with the 3D results. For each mass set, the values of $E$, $L$, and $L_0$ obtained from Eqs. \eqref{ine} and \eqref{inl} are listed in Table \ref{elv}.
\begin{table}[h]
	\caption{\label{elv} Total Energy $(E)$, Total Angular momentum $(L)$, and the three values of $L_0$ for each mass set.}
	\centering
    \resizebox{\linewidth}{!}{
	\begin{tabular}{cccccc} 
		\hline\hline
		Mass set & $E$ & $L$ & $L_0^{(1)}$ & $L_0^{(2)}$ & $L_0^{(3)}$\\
		\hline
		15,15,15 & -1065.92  & 577.15 & 526.86 & 526.86 & 526.86 \\
		12.5,15,17.5 & -1196.69 & 646.92 & 602.01 & 476.67 & 395.08 \\
		12,15,18 & -1222.25 & 660.35 & 616.67 & 462.79 & 371.10 \\
		10,10,20 & -908.00 & 513.02 & 478.39 & 478.39 & 207.15 \\
		10,15,20 & -1322.53 & 712.45 & 674.20 & 396.39 & 282.04 \\
		10,20,20 & -1737.05 & 888.58 & 847.22 & 345.88 & 345.88 \\
		8,21,21 & -1873.64  & 956.05 & 921.58 & 260.77 & 260.77 \\
		5,15,25 & -1559.40 & 833.04 & 811.68 & 180.37 & 102.67 \\
		\hline
	\end{tabular} }
    \tablefoot{Note that $L\geq L_0$. See Eqs. \eqref{ine} and \eqref{inl} and the paragraph below.}
\end{table}

\subsection{Details of the code}

For the purpose of obtaining three-body statistics from simulations, we have developed a numerical code capable of evolving three-body systems in an arbitrary number of dimensions. We employ the fourth-order symplectic algorithm, the Forest--Ruth integrator \citep{1990PhyD...43..105F}. The code records several quantities of interest for further statistical analysis. 

Symplectic integrators do not generally permit variable time stepping, making them less suitable for close encounters, where small time steps are often preferable, and for long excursions, where larger time steps are often preferable. To overcome this limitation, we employ a two-body Keplerian approximation during both of these situations.

When two bodies of the system undergo a sufficiently close and rapid encounter, their relative trajectory is temporarily computed analytically as a two-body system, neglecting the influence of the third body. Over the resulting time interval, the center of mass of this binary and the single body are likewise treated as a two-body system, with their motion computed analytically.

When one body undergoes a long sub-escape excursion, its motion relative to the center of mass of the other two bodies is computed analytically as a two-body system. Simultaneously, the relative motion of the remaining two bodies is computed analytically as an independent two-body system.

The relative error in total energy $(E)$ is monitored periodically at every crossing time $\tau_{\text{cr}}$ for the system \footnote{The crossing time for a system is defined as $\tau_{\text{cr}}:=\frac{GM^{5/2}}{(-2E)^{3/2}}$, where $M$ denotes the total mass.} throughout the simulation to ensure that it is bounded by $10^{-5}$. Whenever this criterion is not satisfied, the simulation is repeated with the time step reduced by a factor of 10. In addition, the relative error in total angular momentum $(L)$ also remains bounded by $10^{-5}$. Due to the chaotic nature of the three-body systems, an individual simulation result would not be reliable for the purpose of determining the exact trajectories, even though the overall errors are small. But the aggregated outcome results obtained after performing 1 million simulations would converge in a statistical sense, enabling the comparison with theoretical predictions.  
%\begin{equation*}
%    \left| \frac{E(t) - E_0}{E_0} \right| \leq 10^{-5}
%\end{equation*}

To detect the breakup of a three-body system, we developed an algorithm that determines whether the system has evolved into a bound binary and an escaping single body. Specifically, the algorithm first checks whether any two bodies form a bound binary whose apoapsis separation is at least ten times smaller than their distance from the single body. If this condition is satisfied, the algorithm then checks whether the center of mass of the bound binary is on a hyperbolic orbit with respect to the single body and moving away from it. If both conditions are met, the system is classified as having undergone breakup. This algorithm is executed periodically throughout the simulation at every crossing time of the system. Tests of the algorithm on randomly generated systems yield fewer than $1$ in $10^5$ false-positive cases, with no false-negative cases observed. We have also implemented an algorithm to detect triple escapes, even though these outcomes do not occur in practice due to negative total energy.

A more detailed description of the code, together with benchmarking results, will be presented in a dedicated paper (E. Springer et al. in prep.).

In addition, we use a previous 3-body code \citep{shoval_thesis}\footnote{The Thesis, including the code, can be accessed at: \href{https://drive.google.com/file/d/1JRlbpF8UDSjpnC2UpKZjaQZOKO_mOsu3/view}{Link}} as well as the TSUNAMI N-body code \citep{2019ApJ...875...42T,2023IAUS..362..404T} to independently verify our simulation results. TSUNAMI performs evolution of the three-body systems using a substantially different numerical approach, combining a leapfrog integrator with Bulirsch--Stoer extrapolation. In addition, the algorithms used to detect various situations in the three-body evolution are also designed, implemented, and tested independently in the two simulation codes.

Although the code described earlier in this subsection is used to generate the simulation data and perform the analyses presented in this paper, we find that the resulting statistical distributions and conclusions remain consistent with those obtained using the TSUNAMI code.

\subsection{Simulation Data}
We record the following quantities after the completion of each simulation: 

{\it Initial Conditions}: Initial positions of the three bodies, Initial velocities of the three bodies, Initial binary phase 

{\it Final state}: Orbital elements of the inner binary, Orbital elements of the outer binary, Angular momentum of the inner binary, Angular momentum of the outer binary, Identity of the escaper

{\it Full interaction}: Lifetime of the system, Time at which the last sub-escape excursion ended, Total time spent in the sub-escape excursions, Scramble number

\section{Simulation results}\label{secres}

In this section, we first describe the two conditions (cuts) used to select the appropriate subsets of simulations. We then present the simulation results and perform further analysis for various statistical quantities. Next, the comparisons with the theoretical predictions are made. For some quantities we find agreement with theoretical expectations, while for others we observe deviations.

\subsection{Non-prompt ejections cut and the Ergodic cut}\label{lecs}

The Flux-based theory predicts statistical quantities under the assumption that the disintegration originates from the ergodic region of phase space. In the context of simulations, this implies that the final escape occurs immediately following an ergodic motion. Here, an ergodic motion is considered to consist of at least 3--4 scramble states, such that information about the preceding state has effectively been lost.

To compare the simulation results with theoretical predictions, appropriate subsets of simulations must be selected. First, we remove the subset of simulations corresponding to {\it prompt ejections}, in which an immediate escape follows the initial in-fall of the binary and the single body, since such interactions exhibit non-chaotic behavior. This is implemented using a lifetime cut, where only the systems with lifetimes $\tau_D\geq30$ yrs are considered, since the in-fall time is approximately $27$ yrs. We refer to this condition as the {\it non-prompt ejections cut}.

Next, we also require that the escape originates from an ergodic region of phase space. This condition is imposed by selecting systems with $\tau_{\text{gap}}\geq10$ yrs, where $\tau_{\text{gap}}$ denotes the time interval between the end of the last sub-escape excursion and the final escape. This ensures that the system underwent an ergodic motion before escape, consisting of at least 3--4 scramble states, given that each scramble state typically lasts $2$--$3$ yrs, as observed from the simulations. 

Additionally, we impose the condition $\tau_D\geq80$ yrs based on the observation that interactions with $\tau_D<80$ yrs exhibit band-like structures in the initial-condition phase-space maps, indicating non-chaotic behavior. We refer to the combined conditions $\tau_D\geq80$ yrs and $\tau_{\text{gap}}\geq10$ yrs as the {\it ergodic cut}.

In summary,
\begin{equation}
	\begin{split}
		\text{non-prompt ejections cut:}&\quad \tau_D\geq30\text{yrs} \\
		\text{ergodic cut:}&\quad \tau_D\geq 80\text{yrs} \text{ and } \tau_{\text{gap}}\geq10\text{yrs}
	\end{split}
\end{equation}
We choose the subset of simulations satisfying the ergodic cut condition in order to compare the results with theoretical predictions. Making the ergodic cut more restrictive does not appreciably alter the statistical results, indicating that the chosen cut is convergent.

\subsection{Escape probabilities}\label{epds}

We select the subset of simulations satisfying the ergodic cut. The escape probabilities obtained from this subset are presented in figure \ref{epfig}. In the same figure, we compare these results with the predictions derived from the flux-based theory in Section \ref{pes} (the planar prediction in Eq. \eqref{ep2} and the 3D prediction in Eq. \eqref{ep3}), as well as with alternative theoretical derivations of \citep{2006tbp..book.....V}, \citep{2019Natur.576..406S}, \citep{2021PhRvX..11c1020G}. The numerical values corresponding to Figure \ref{epfig} are listed in Table \ref{ept} in Appendix \ref{resapp}, where the numerical errors $(1\sigma)$ associated with simulation results $P_{s}^{(2d)}$ are also shown.

We find that the escape probabilities obtained from the planar simulations using the ergodic cut agree with the corresponding 3D simulation results (presented in Table 2 of MKTL21), as well as the 3D theoretical predictions \eqref{ep3} of Flux-based theory, at approximately the $1\%$ level. Consequently, they are in clear disagreement with the planar prediction of the Flux-based theory under the emissivity-blind approximation, as well as other theoretical predictions.

\begin{figure}[ht!]
	\centering
	\includegraphics[width=\hsize]{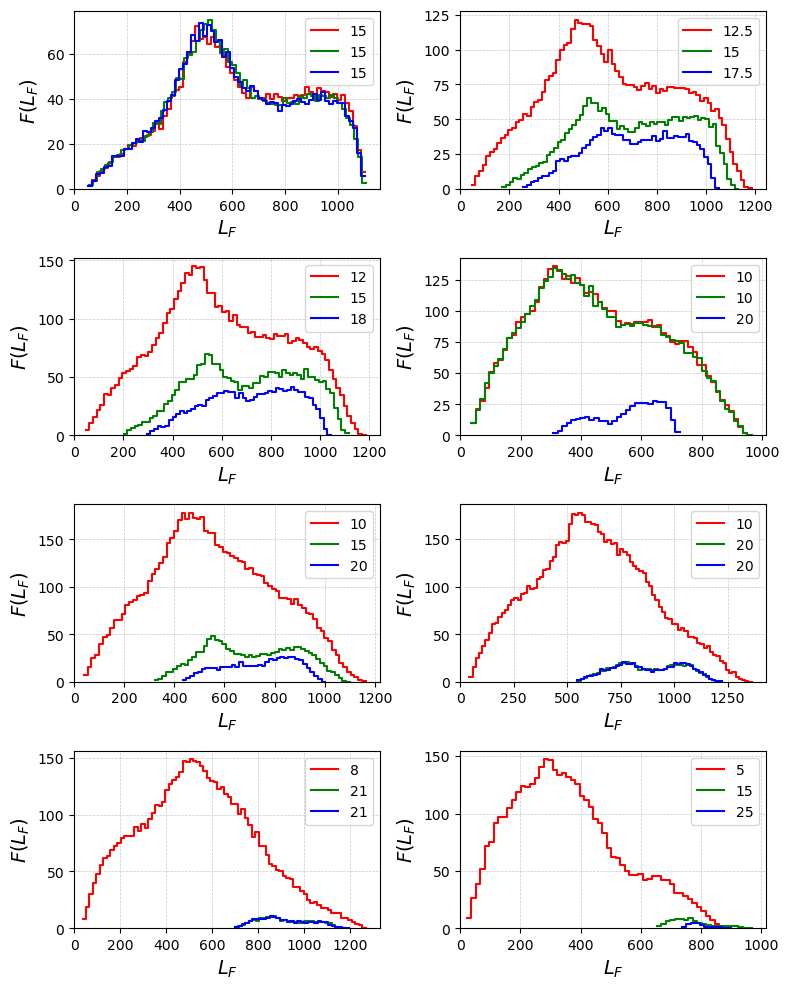}
	\caption{Distribution of the angular momentum of the remnant outer binary ($L_F$) for the planar problem. $F(L_F)$ denotes the frequency statistic used in these plots, defined as the number of observations divided by the bin width.}
	\label{damo}
\end{figure}

\subsection{Marginal escapes}\label{meds}
In this subsection, we compare the distribution of the angular momentum of the remnant outer binary ($L_F$) with the theoretical predictions presented in Section \ref{mes} for the case of marginal escapes. The $L_F$ distributions for simulations satisfying the ergodic cut are shown in Figure \ref{damo}. Marginal escapes correspond to the leftmost regions of these distributions. 

\begin{figure}[ht!]
	\centering
	\includegraphics[width=\hsize]{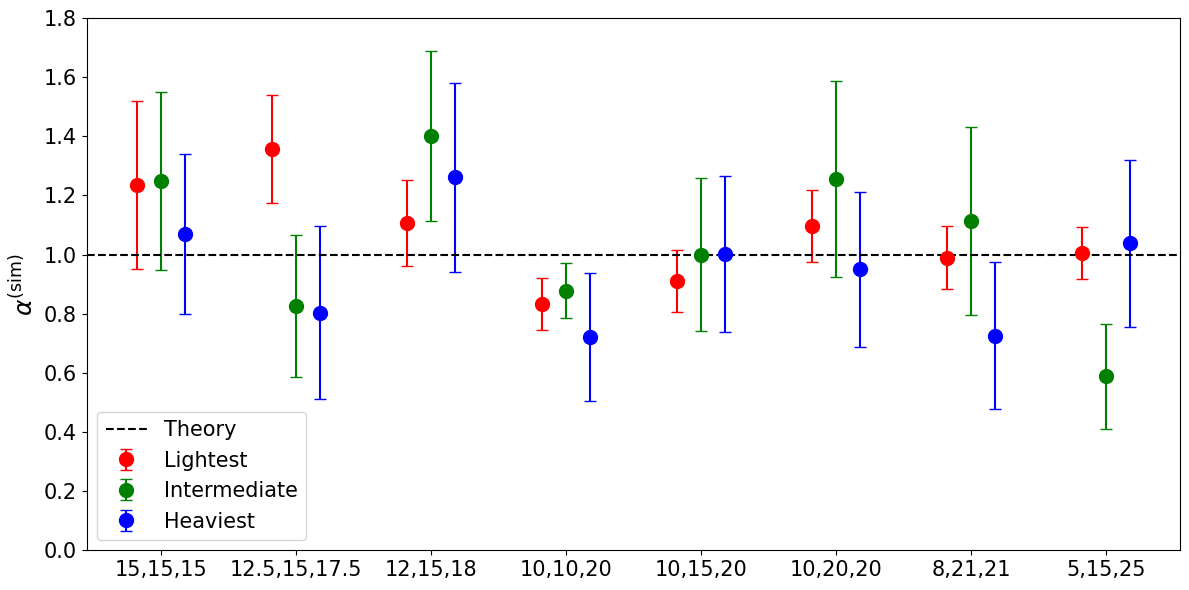}
	\caption{Maximum Likelihood Estimation values and standard errors of $\alpha^{(sim)}$ in \eqref{lf2de} for all the mass sets. The theoretical prediction of $\alpha=1$ is shown as a dashed black line.}
	\label{msfig}
\end{figure}

 The predicted distribution given by Eq. \eqref{lf2d} is 
\begin{equation}\label{lf2de}
	dP_s \propto (L_F-L_{F,c})^{\alpha} dL_F
\end{equation}
where $\alpha=1$. To quantify the level of agreement between the distribution obtained from simulations with the theoretical prediction, we estimate the values and standard errors of the power-law index $\alpha$ using Maximum Likelihood Estimation technique and present the results in figure \ref{msfig}. In this estimation, we consider the initial part of the $L_F$ distribution with the condition $L_{F,c}\leq L_F \leq L_{F,c}+30$. Actual values of the estimates that the Figure \ref{msfig} is based on are listed in Table \ref{mlet} in Appendix \ref{resapp}. The table also lists the values of $L_{F,c}$ measured from the simulations.

Using the inverse-variance weighting method, we combine the results from all 24 cases listed in Table \ref{mlet}. The resulting aggregated value and standard error of the parameter $\alpha$ is
\begin{equation}
    \alpha = 0.964 \pm 0.033
\end{equation}
This value is in good agreement with the theoretical prediction.

%Since the predicted distribution is linear, it is useful to present the distribution obtained from the simulations in a log-log plot and perform linear regression analysis. 

%However, the linear regression analysis needs to be performed carefully. The distribution is expected to be linear only in the initial part near the threshold value, $L_{F,c}$. So, we need to perform regression over a range bounded by a maximum value, $L_F^{(\text{max})}$. Also, we expect that there would be statistical fluctuations in the initial part of the distribution, due to very small number of systems falling within the bins in the initial part of the distribution. So we need to further restrict the range for the regression by a minimum value, $L_F^{(\text{min})}$. In conclusion, we perform linear regression over the range $L_F^{(\text{min})}\leq L_F \leq L_F^{(\text{max})}$.

%The distributions of $L_F$ are restricted to be within the range $L-L_0\leq L_F \leq L+L_0$. So we find it natural to set $L_F^{(\text{max})} = 0.2~ L_0$. We set $L_F^{(\text{min})} = 10$ for convenience. 

\subsection{Eccentricities}\label{ioes}

In this subsection, we compare the eccentricity distributions of the remnant inner and outer binaries with the theoretical predictions of Eq. \eqref{ibe} and \eqref{obe} presented in Section \ref{eds}.

We transform the $e_B$ distributions obtained from the simulations such that the transformed count ($T$) in each bin is
\begin{equation}\label{trct}
    T = \frac{C}{\sqrt{1+C^2}},
\end{equation}
where $C$ denotes the original count. 
It is obtained from \eqref{ibe} by replacing $dP_s^{(2d)}/de_B$ with $C$, normalizing the distribution (the normalization factor equal to unity), inverting the relation $C=C(e_B)$ to express $e_B$ in terms of the predicted $C$, and finally replacing $e_B$ with $T$, so $T$ is predicted to equal $e_B$. 
This transformation facilitates comparison between the simulation results and the theoretical prediction by converting the theoretical distribution into a linear relation. To quantify the level of agreement, we use the coefficient of determination, $R^2$.

\begin{figure}[ht!]
	\centering
	\includegraphics[width=\hsize]{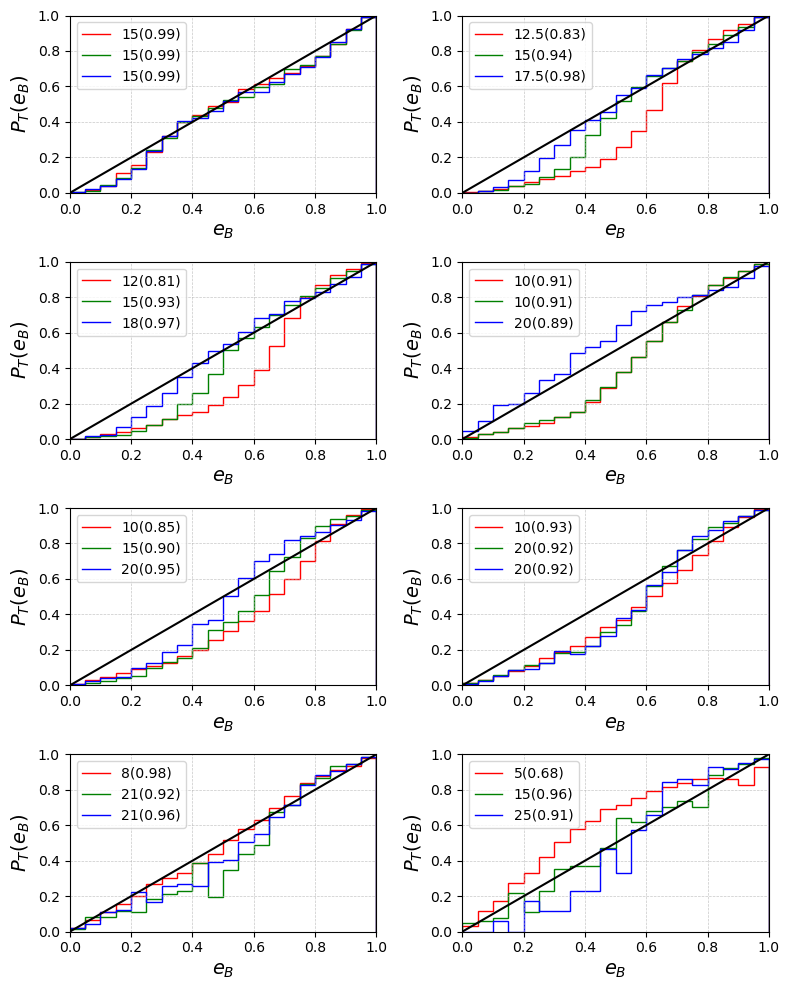}
    \caption{
    Distribution of the eccentricity of the remnant inner binary for the planar problem, where the bin counts are transformed by the transformation \eqref{trct}. Theoretical prediction is shown by the black line. The $R^2$ values quantifying the agreement between the simulation distribution and the theoretical prediction are shown in parentheses corresponding to each distribution.}
	\label{iec}
\end{figure}

The transformed $e_B$ distributions obtained from simulations satisfying the ergodic cut are shown in Figure \ref{iec}. The corresponding $R^2$ values are given in parentheses for each distribution. Overall, the emissivity-blind approximation reproduces the simulated distributions reasonably well, although systematic deviations are visible in certain regions.

\begin{figure}[h]
	\centering
	\includegraphics[width=\hsize]{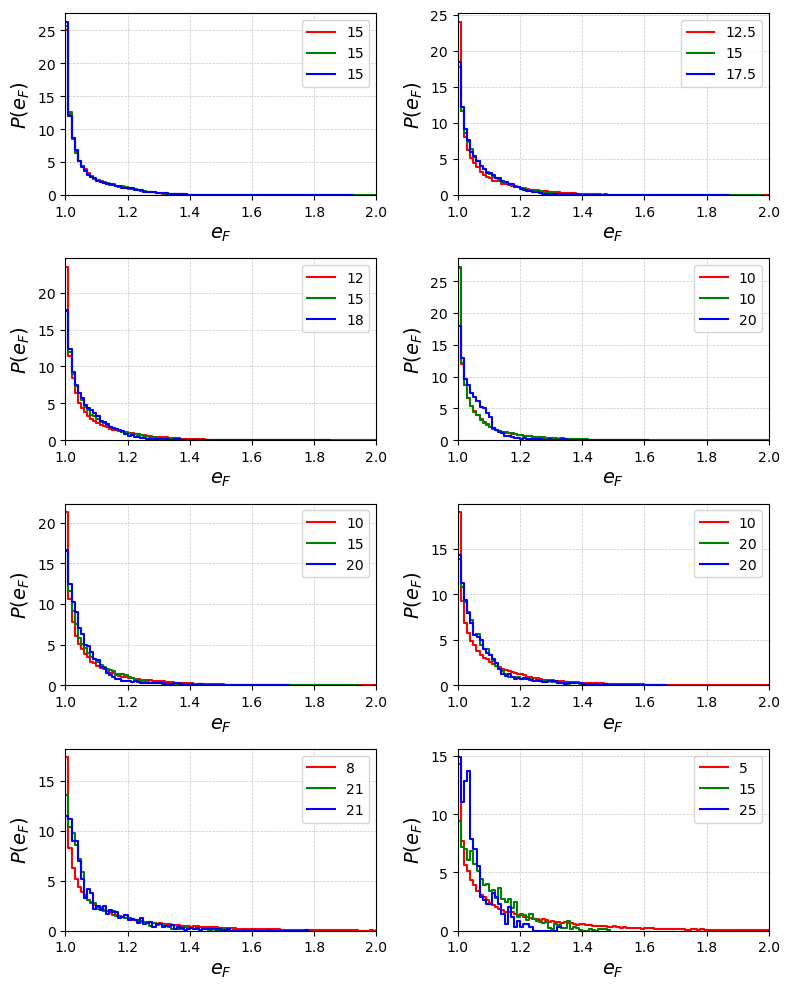}
	\caption{
    Distribution of the eccentricity of the remnant outer binary for the planar problem. Although $1\leq e_F\leq \infty$, the horizontal axis is restricted to the range $1\leq e_F\leq2$ for clarity.}
	\label{oec}
\end{figure}

The eccentricity distributions of the remnant outer binaries obtained from simulations satisfying the ergodic cut are shown in Figure \ref{oec}. The distributions decay rapidly for large $e_F$, as expected from the theoretical derivation of a power-law decay. Further quantitative analysis of the large $e_F$ region of the distribution is prevented by an insufficient number of systems with large $e_F$ values.

%As discussed in Section \ref{eds}, the theoretical prediction derived under the emissivity-blind approximation is non-normalizable and therefore does not appear to be physically meaningful. Consequently, a detailed comparison between the theoretical prediction and the simulated distribution is deferred to future work.

\subsection{Lifetime distributions}\label{ldds}
\begin{figure}[h]
    \centering
    \includegraphics[width=\hsize]{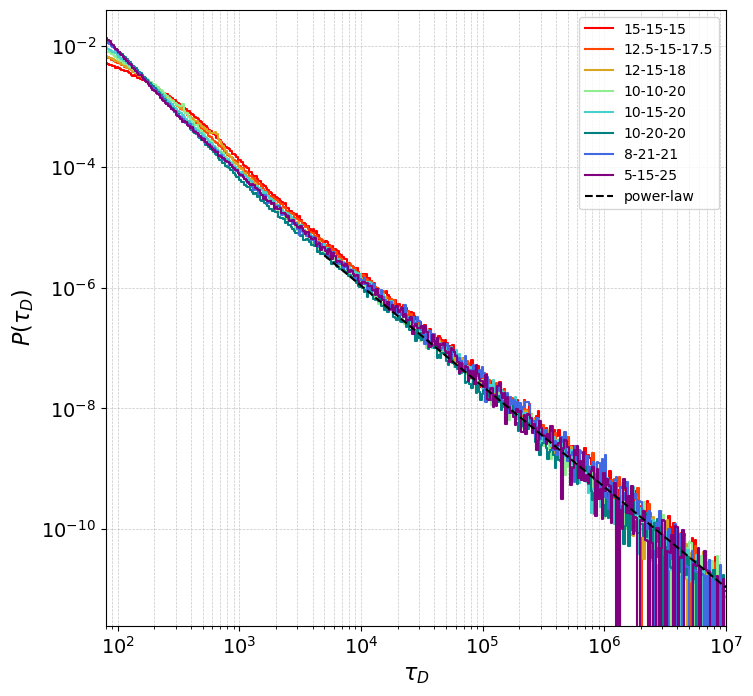}
    \caption{Differential lifetime distributions in log-log scale for all mass sets with lifetimes $\tau_D\geq80$ yrs. The theoretically expected power-law behavior $(\tau_D^{-5/3})$ is shown as a black dashed line.}
    \label{dlf}
\end{figure}

\begin{figure*}[h]
	\centering
	\includegraphics[width=0.8\linewidth]{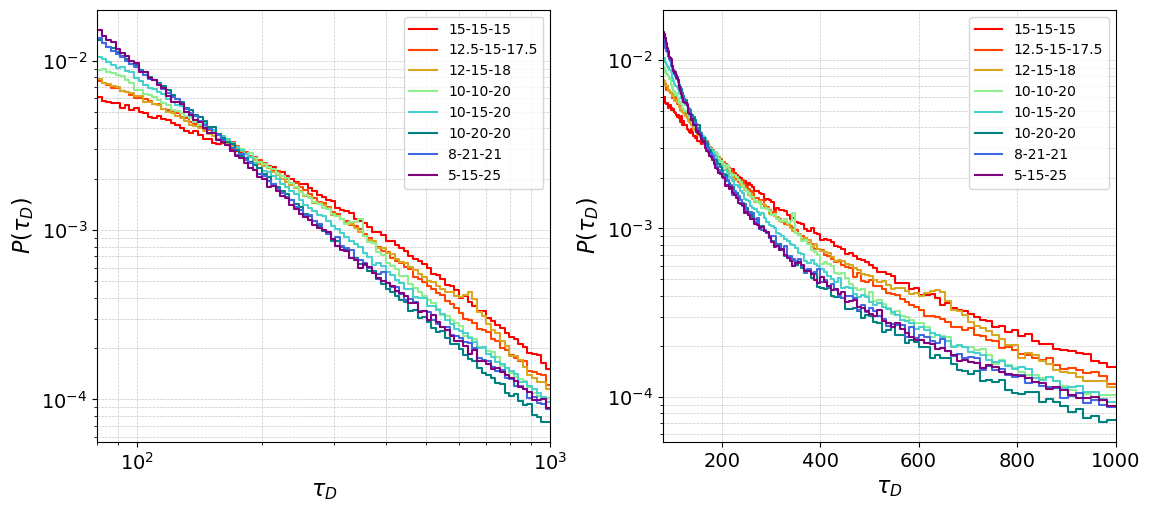}
	\caption{Differential lifetime distributions in log-log and semi-log scales for systems with lifetimes in the range $80\leq\tau_D\leq1000$ yrs. This figure presents a zoomed-in version of Figure \ref{dlf}.}
	\label{dls}
\end{figure*}

We consider the lifetime distributions for simulations with lifetimes $\tau_D\geq80$ yrs. Differential lifetime distributions are plotted in the log-log scale in Figure \ref{dlf}, along with the theoretically expected power-law with index $-5/3$ for large lifetimes. The distributions have large fluctuations for extremely long lifetimes due to the expected scarcity of systems in this region. The linear regression fits for the power-law index of the lifetime distributions in the region $3000\leq\tau_D\leq10^6$ using log-log scale are given in Table \ref{ple}. We find that these observations are in agreement with theoretical predictions discussed in section \ref{lds}. \footnote{The corresponding results from 3D simulations are presented in Table 5 of MKTL21.}
\begin{table}[h]
    \caption{ \label{ple} Power-law index fits for large lifetimes. }
	\centering
	\begin{tabular}{cc} 
		\hline\hline
		Mass Set & Power-law index\\
		\hline
		15,15,15 & $-1.692 \pm 0.015$ \\
		12.5,15,17.5 & $-1.704 \pm 0.014$  \\
		12,15,18 & $-1.703 \pm 0.014$ \\
		10,10,20 & $-1.693 \pm 0.014$ \\
		10,15,20 & $-1.721 \pm 0.019$ \\
		10,20,20 & $-1.698 \pm 0.018$\\
		8,21,21 & $-1.651 \pm 0.021$ \\
		5,15,25 & $-1.708 \pm 0.021$ \\
		\hline
	\end{tabular}
	\tablefoot{Power-law indices obtained by performing linear regression on the lifetime distributions in log-log space for the region $3000\leq\tau_D\leq10^6$. Kepler's third law predicts a value of $-5/3$.}
\end{table}
In Figure \ref{dls} we present a zoomed-in version of Figure \ref{dlf} in log-log and semi-log scales to focus on the initial part of the distribution. From the semi-log plot, it is clear that the initial part of the lifetime distribution is not purely exponential (since it does not appear linear in the semi-log plot), indicating the mixing phenomenon between the exponential decay and the power-law decay. The log-log scale plot shows the {\it bump} feature that is clearly visible for the mass sets with low mass contrast vanishes as the mass contrast increases. This bump feature is an indication of the exponential decay. The vanishing of the bump feature implies that as the mass contrast increases, the system tends to undergo an increasing number of sub-escape excursions rather than chaotic motion. Hence, the exponential decay is suppressed, and the power-law decay becomes more dominant. 

\begin{figure}[h]
	\centering
	\includegraphics[width=\hsize]{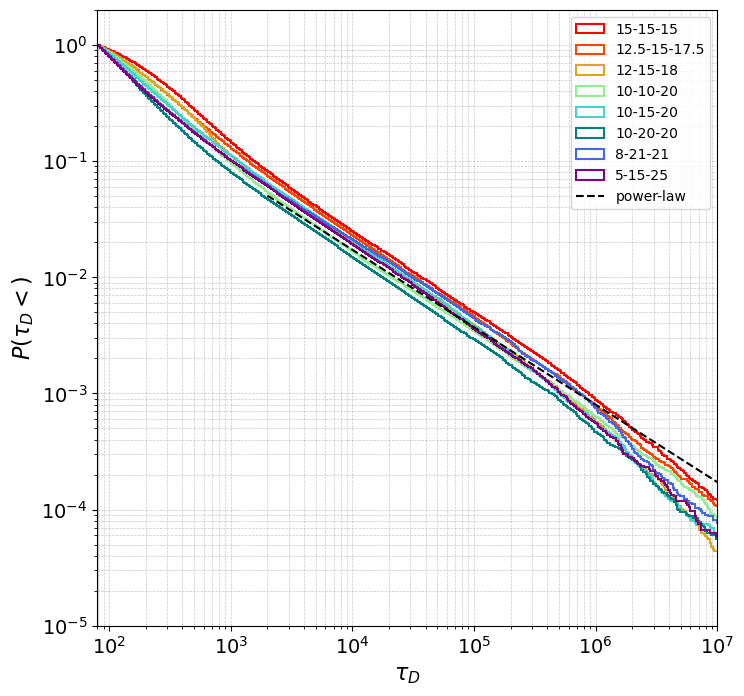}
	\caption{
    Cumulative lifetime distributions  
    (for lifetimes longer than $\tau_D$) 
    in log-log scale for systems with lifetimes $\tau_D\geq80$ yrs. The theoretically expected power-law behavior $(\tau_D^{-2/3})$ is shown as a black dashed line.}
	\label{clf}
\end{figure}

\begin{figure*}[h]
	\centering
	\includegraphics[width=0.8\hsize]{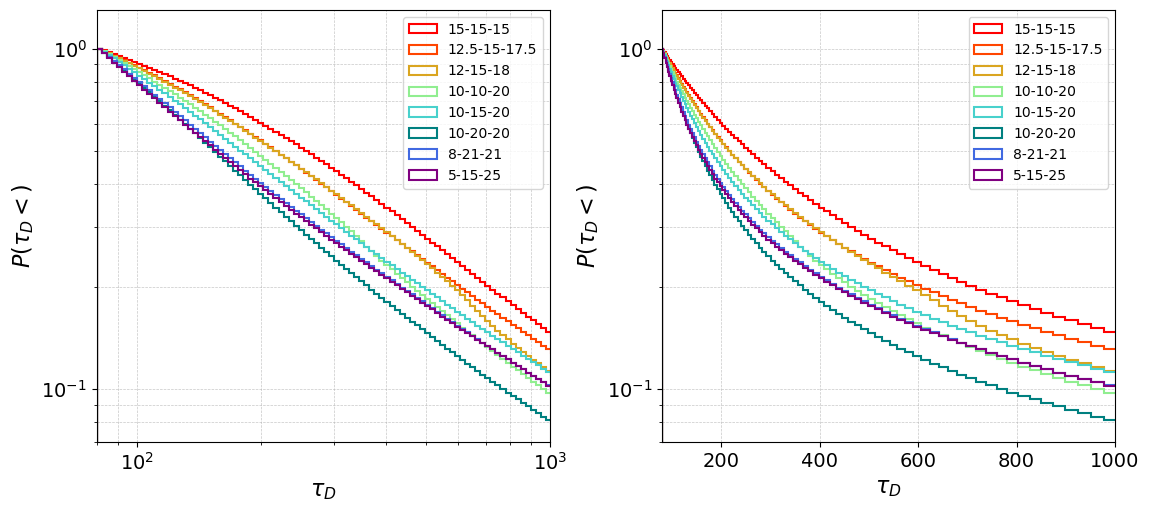}
	\caption{
    Cumulative lifetime distributions in log-log and semi-log scales for systems with lifetimes in the range $80\leq\tau_D\leq1000$ yrs.  This figure presents a zoomed-in version of Figure \ref{clf}.}
	\label{cls}
\end{figure*}

Cumulative lifetime distributions are plotted in Figure \ref{clf}, along with the theoretically expected power-law decay with index $-2/3$. As expected, it can be observed that cumulative distribution is much smoother than the differential distribution, especially for the large lifetimes. In Figure \ref{cls} we present zoomed-in version of Figure \ref{clf} in log-log and semi-log scales to focus on the initial part of the distribution. %We again observe the vanishing of the bump feature when the mass contrast is increased.



The differential lifetime distributions for the planar problem are compared with their three-dimensional counterparts. For all unequal-mass systems, the planar distributions exceed the three-dimensional distributions at early times and fall below them at late times. The equal-mass case exhibits the opposite behavior. To quantify the similarity between the planar and three-dimensional lifetime distributions, we use the overlap coefficient ($0\leq C_O\leq1$),
\begin{equation}\label{ovl}
    C_O := \int \min (P^{(3d)}(\tau_D),P^{(2d)}(\tau_D)) ~d\tau_D 
\end{equation}
where $P^{(3d)}(\tau_D)$ and $P^{(2d)}(\tau_D)$ denote the lifetime distributions for the three-dimensional and planar problems, respectively. The overlap coefficients for all mass sets are listed in Table \ref{ovlt}.
\begin{table}[h]
	\caption{\label{ovlt} Overlap coefficient Eq. \eqref{ovl} between the planar and 3D lifetime distributions.}
	\centering
	\begin{tabular}{cc} 
		\hline\hline
		Mass Set & Overlap Coefficient\\
         & (3D-2D)\\
		\hline
		15,15,15 &  0.98\\
		12.5,15,17.5 &  0.98 \\
		12,15,18 &  0.98\\
		10,10,20 &  0.97\\
		10,15,20 &  0.95\\
		10,20,20 & 0.91\\
		8,21,21 &  0.95\\
		5,15,25 &  0.95\\
		\hline
	\end{tabular}
\end{table}

\subsection{Sub-escape fraction distribution}\label{sefs}

\begin{figure*}[h]
    \centering
    \includegraphics[width=0.75\hsize]{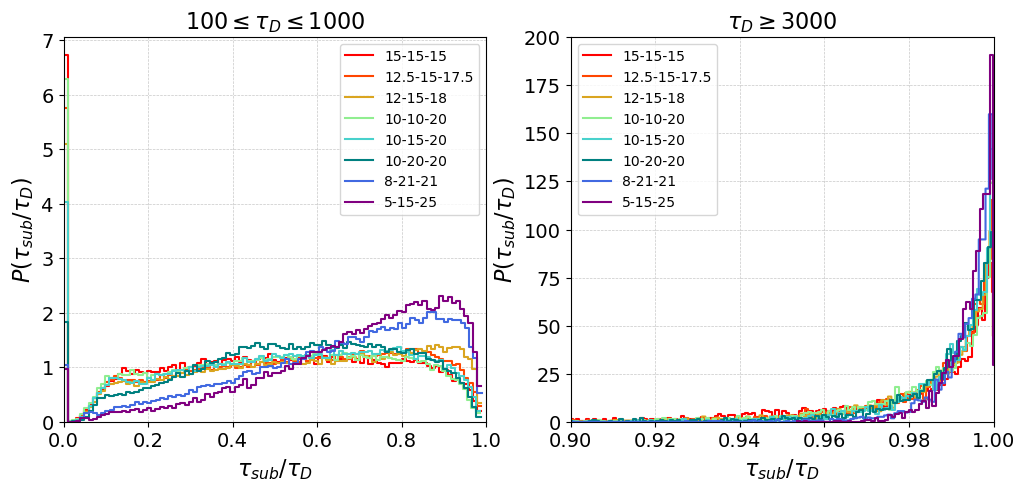}
    \caption{Distributions of the sub-escape fraction for shorter lifetimes ($100\,\mathrm{yrs}\leq\tau_D\leq1000\,\mathrm{yrs}$) and longer lifetimes ($\tau_D\geq3000\,\mathrm{yrs}$). The distributions are constructed using systems satisfying the ergodic cut.}
    \label{sef}
\end{figure*}

We define the sub-escape fraction as the ratio of total time spent in sub-escape excursions to the lifetime of the three-body interaction. Distributions of sub-escape fraction are shown in Figure \ref{sef}, where we have used the subset of simulations satisfying the ergodic cut. For shorter lifetimes, the distribution of sub-escape fractions indicate that both the chaotic motion and the sub-escape excursions contribute substantially, which leads to the mixing effects between the exponential decay and the power law decay. Next, the longer lifetimes are indeed dominated by sub-escape excursions, leading to an almost pure power-law decay as expected. Corresponding results for the 3D case are shown in Fig. 8 of MKTL21, and are qualitatively similar to the planar case.

\subsection{Prograde and Retrograde escapes}\label{prss}

To study the statistics of prograde and retrograde escapes, we consider the systems that satisfy the ergodic cut. We define the Pro/Ret Ratio as the ratio of the number of systems undergoing prograde escapes to those undergoing retrograde escapes. The simulations show that this ratio increases systematically with increasing mass contrast. In contrast, the Flux-based theory predicts a value of unity, independent of the masses and conserved quantities, as discussed in Section \ref{prs}. The simulations also show that the escape probabilities for prograde and retrograde escapes differ, whereas the theory predicts them to be identical. These results are presented in Figure \ref{prfig}. The numerical values underlying Figure \ref{prfig} are listed in Table \ref{prp} in Appendix \ref{resapp}.

\begin{figure*}[h]
    \centering
    \sidecaption
    \includegraphics[width=0.7\hsize]{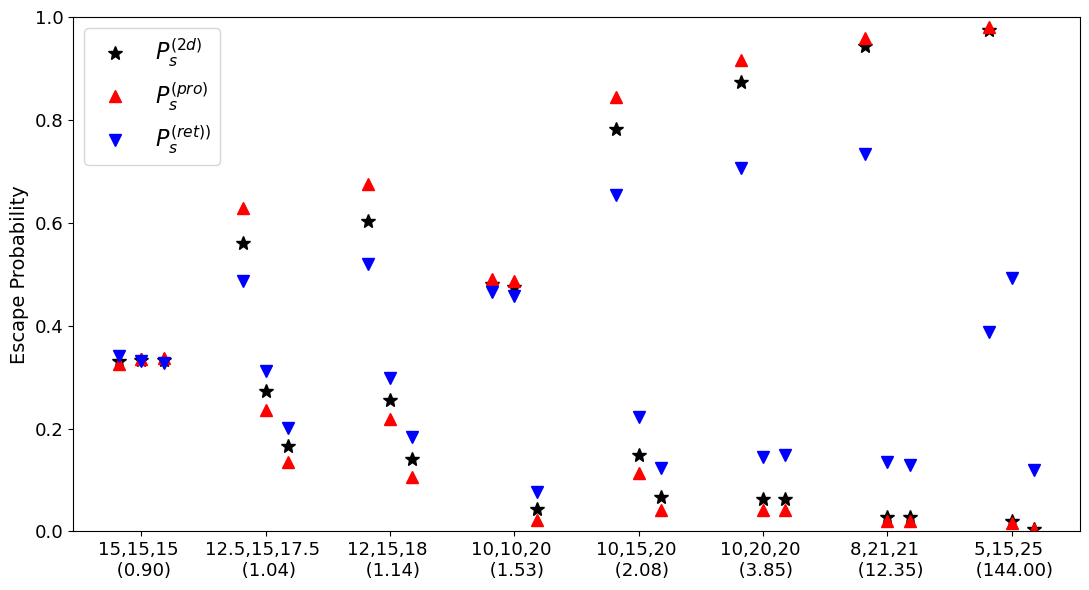}
    \caption{
    The figure shows the overall escape probabilities measured from the simulations ($P_s^{(2d)}$) together with those computed separately for the prograde ($P_s^{(pro)}$) and retrograde ($P_s^{(ret)}$) subsets. For each mass set, the escape probabilities are ordered according to the lightest, intermediate, and heaviest body. The Pro/Ret Ratio, shown in parentheses below each mass set, denotes the ratio of systems undergoing prograde escapes to those undergoing retrograde escapes.\\
    \vspace{13pt}
    }
    \label{prfig}
\end{figure*}

The similarity between the lifetime distributions of prograde and retrograde escapes is quantified using the overlap coefficient ($0\leq C_O\leq1$),
\begin{equation}\label{ovlr}
    C_O := \int \min (P^{(pro)}(\tau_D),P^{(ret)}(\tau_D)) ~d\tau_D 
\end{equation}
where $P^{(pro)}(\tau_D)$ and $P^{(ret)}(\tau_D)$ denote the lifetime distributions for prograde and retrograde escapes, respectively. The overlap coefficients for all mass sets are listed in Table \ref{ovlrt}.
\begin{table}[h]
    \caption{\label{ovlrt} Overlap coefficient \eqref{ovlr} between the lifetime distributions for the prograde and retrograde escapes.}
	\centering
	\begin{tabular}{cc} 
		\hline\hline
		Mass Set & Overlap Coefficient \\
         & (prograde-retrograde) \\
		\hline
		15,15,15 &  0.98\\
		12.5,15,17.5 &  0.98 \\
		12,15,18 &  0.98\\
		10,10,20 &  0.96\\
		10,15,20 &  0.97\\
		10,20,20 & 0.91\\
		8,21,21 &  0.89\\
		5,15,25 &  0.83\\
		\hline
	\end{tabular}
\end{table}

%The lifetime distributions for the prograde escapes are found to be similar to the lifetime distributions for the retrograde escapes, where the percentage difference is typically bounded by $40\%$.

%The percentage difference between the planar and 3D differential distributions is within $20\%$, except for the mass set $20,20,10$ where the difference is within $40\%$. {\color{red}(These results could be presented in a better way.)}

\section{Summary and Future directions}\label{secsum}

In this paper, we studied the non-hierarchical planar three-body problem with negative total energy using a statistical approach. We investigated the problem from both theoretical and numerical perspectives.

Our primary focus was the Flux-based statistical theory introduced in \citep{2021CeMDA.133...17K}, which relates the decay rate of three-body systems to the flux of phase-space volume, as summarized in section \ref{thint}. For the planar problem, we derived a closed-form expression for the total flux of the phase-space volume in section \ref{dfs}, which serves as the basis for theoretical predictions of various statistical outcomes. The lack of a closed-form expression for the emissivity was circumvented by concentrating on quantities for which the effects of emissivity can reasonably be neglected. Using this framework, we derived theoretical predictions for: (a) the escape probabilities of the three bodies in section \ref{pes}, (b) the distribution of the angular momentum of the remnant outer binary in the case of marginal escapes in section \ref{mes}, (c) the eccentricity distributions of the remnant inner and outer binaries in section \ref{eds}, (d) the lifetime distribution in section \ref{lds} and (e) the relative prevalence of prograde and retrograde escapes in section \ref{prs}.

As discussed in section \ref{secsim}, we then performed one million three-body simulations with randomized initial conditions to measure the corresponding statistical quantities. A summary of the comparison between the simulations and the theoretical predictions is as follows:

The escape probabilities measured from the simulations were found to disagree with the prediction derived from the Flux-based theory for the planar problem. Surprisingly, however, they agree at the one-percent level with the predictions from the Flux-based theory as well as the simulation results for the corresponding 3D problem. This was presented in section \ref{epds}.

The angular-momentum distribution of the remnant outer binary for marginal escapes was measured from the simulations. A Maximum Likelihood Estimation and aggregation analysis of the data showed good agreement with the theoretical prediction. This was presented in section \ref{meds}.

The eccentricity distribution of the remnant inner binary was found to agree remarkably well with the theoretical prediction, despite the lack of a clear justification for the emissivity-blind approximation. The eccentricity distribution of the remnant outer binary was also measured from the simulations and compared with theoretical prediction. This was presented in section \ref{ioes}.

The lifetime distributions were measured from the simulations. The power-law index of the late-time portion of the distribution was found to be consistent with the theoretical prediction. The early-time behavior of the lifetime distribution was also found to agree with theoretical expectations. This was presented in section \ref{ldds}. A complete theoretical description of the lifetime distribution requires the consideration of, among other ingredients, the total phase-space volume derived in \citep{2022CeMDA.134...55D,2025CeMDA.137...32D}. We plan to revisit this problem in future work.

The distribution of the sub-escape fraction was measured from the simulations. The results provide additional evidence for the coexistence of ergodic motion and sub-escape excursions during the early stages of the lifetime distribution. This was presented in section \ref{sefs}.

Finally, the relative prevalence of prograde and retrograde escapes observed in the simulations was found to deviate from the theoretical prediction. This was presented in section \ref{prss}.

A possible explanation for the discrepancies in the escape probabilities and the relative prevalence of prograde and retrograde escapes is the breakdown of the emissivity-blind approximation. Although this approximation appears to hold well in the three-dimensional case studied in MKTL21 \citep{2021MNRAS.506..692M}, it may not remain valid for the planar problem. In addition, the assumption of probability equilibration, and consequently the applicability of the statistical approach itself, deserves further investigation in the planar case because of its reduced number of degrees of freedom relative to the three-dimensional system. We plan to examine both of these issues in future work.

Another natural extension of the present work is to investigate the implications of our results for specific astrophysical scenarios, such as the formation of planets in protoplanetary disks and triple black-hole interactions in AGN disks, and to develop statistical descriptions tailored to these environments.

Other possible directions for future research include: (a) performing simulations in the regime $L\leq L_0$ and comparing the results with theoretical predictions, and (b) extending the analysis to general power-law interaction potentials rather than restricting it to Newtonian gravity to test the robustness of the predictive power of statistical theories.

\FloatBarrier

\begin{acknowledgements}
Y.~D.~has been supported
    by the Israel Science Foundation (ISF) grant No. 1698/22,
    and by the US-Israel Binational Science Fund (BSF) grant No. 2024816, matching
    NSF-BSF award No. 2513321.
S.~Z.~has been supported
    by the Israel Science Foundation (ISF) grant No. 1698/22,
    by the US-Israel Binational Science Fund (BSF) grant No. 2020245,
    and by O.~B.'s Alon scholarship to young faculty,
    provided by the Israeli Council for Higher Education's Planning and Budgeting Committee.
    B.~K.~has been supported by the Israel Science Foundation (grant no. 2058/25) and by a grant from Israel’s Council of Higher Education.
    We want to thank Yoav Zak and Daniel Rozental for helpful discussions.
\end{acknowledgements}

\bibliographystyle{aa} 
\bibliography{ptb}

\begin{appendix}

\begin{table*}[h]
\section{List of symbols}\label{symsec}
	\centering
	\begin{tabular}{cc} 
		\hline\hline
		Symbol & Definition \\
		\hline
		$E,L$ & Total Energy and Angular momentum for the three-body system\\
		$F$ & Total flux of phase-space volume\\
		$\bar\sigma$ & Regularized volume of phase-space \\
		$E_B,E_F$ & Total Energy of inner and outer binary $(E_B+E_F=E)$ \\
		$L_B,L_F$ & Angular momentum of inner and outer binary $(L_B+L_F=L)$ \\
		$L_0$ & Maximum value of %for the 
        $|L_B|$ \\
		$L_{F,c}$ & Minimum value of $L_F$ if $L\geq L_0$ $(L_{F,c}=L-L_0)$\\
		$\phi_B, \phi_F$ & Pericenter angle for inner and outer binary \\
	    $\mu_B,\mu_F$ & Reduced mass of inner and outer binary \\
	    $\alpha_B,\alpha_F$ & Potential strength constant of inner and outer binary \\
	    $k_B,k_F$ & Binary constants: $\mu_B\alpha_B^2, ~\mu_F\alpha_F^2$ \\
	    $e_B, e_F$ & Eccentricity of inner and outer binary\\
	    $\tau_D$ & Lifetime of a three-body system\\
	    $\tau_{gap}$ & Time difference between the last sub-escape excursion and the final escape\\
	    $\tau_{sub}$ & Total time spent in sub-escape excursions for a three-body system\\
	    ${s,a,b}$ & $s$: escaper, $a,b$: inner binary\\
	    $P_s$ & Probability of escape for body $s$\\
		\hline
	\end{tabular}
	\caption{Definitions for the list of symbols frequently used}
	\label{symb}
\end{table*}

%\begin{landscape}
\begin{table*}[h]
\section{Tables of Results}\label{resapp}
	\centering
	\small
\begin{tabular}{ccccccccc}
\hline\hline
Mass set & Escaper & \%(np) & $P_{s}^{(2d,np)}$ & \%(es) & $P_{s}^{(2d)}$ & Kol 3D & Kol 2D & VK 2D \\
\hline
& 15 & & $0.3389 \pm 0.0007$ & & $0.3339 \pm 0.0014$ & 0.3333 & 0.3333 & 0.3333 \\
15,15,15 & 15 & 46\% & $0.3309 \pm 0.0007$ & 12\% & $0.3337 \pm 0.0014$ & 0.3333 & 0.3333 & 0.3333 \\
& 15 & & $0.3303 \pm 0.0007$ & & $0.3324 \pm 0.0014$ & 0.3333 & 0.3333 & 0.3333 \\
\hline
& 12.5 & & $0.5653 \pm 0.0006$ & & $0.5589 \pm 0.0014$ & 0.5621 & 0.4860 & 0.5146 \\
12.5,15,17.5 & 15 & 66\% & $0.2634 \pm 0.0005$ & 13\% & $0.2734 \pm 0.0012$ & 0.2790 & 0.3047 & 0.2978 \\
& 17.5 & & $0.1714 \pm 0.0005$ & & $0.1677 \pm 0.0010$ & 0.1589 & 0.2093 & 0.1875 \\
\hline
& 12 & & $0.5925 \pm 0.0006$ & & $0.6028 \pm 0.0013$ & 0.6095 & 0.5194 & 0.5530 \\
12,15,18 & 15 & 74\% & $0.2482 \pm 0.0005$ & 14\% & $0.2555 \pm 0.0012$ & 0.2576 & 0.2925 & 0.2831 \\
& 18 & & $0.1593 \pm 0.0004$ & & $0.1418 \pm 0.0009$ & 0.1328 & 0.1881 & 0.1639 \\
\hline
& 10 & & $0.4949 \pm 0.0006$ & & $0.4811 \pm 0.0013$ & 0.4805 & 0.4571 & 0.4706 \\
10,10,20 & 10 & 78\% & $0.4494 \pm 0.0006$ & 15\% & $0.4748 \pm 0.0013$ & 0.4805 & 0.4571 & 0.4706 \\
& 20 & & $0.0556 \pm 0.0003$ & & $0.0442 \pm 0.0005$ & 0.0390 & 0.0857 & 0.0588 \\
\hline
& 10 & & $0.7576 \pm 0.0004$ & & $0.7827 \pm 0.0011$ & 0.7834 & 0.6576 & 0.7036 \\
10,15,20 & 15 & 91\% & $0.1652 \pm 0.0004$ & 13\% & $0.1491 \pm 0.0010$ & 0.1592 & 0.2273 & 0.2085 \\
& 20 & & $0.0772 \pm 0.0003$ & & $0.0682 \pm 0.0007$ & 0.0573 & 0.1151 & 0.0879 \\
\hline
& 10 & & $0.8173 \pm 0.0004$ & & $0.8735 \pm 0.0009$ & 0.8802 & 0.7500 & 0.8000 \\
10,20,20 & 20 & 99\% & $0.0911 \pm 0.0003$ & 13\% & $0.0632 \pm 0.0007$ & 0.0599 & 0.1250 & 0.1000 \\
& 20 & & $0.0915 \pm 0.0003$ & & $0.0633 \pm 0.0007$ & 0.0599 & 0.1250 & 0.1000 \\
\hline
& 8 & & $0.9112 \pm 0.0003$ & & $0.9428 \pm 0.0008$ & 0.9567 & 0.8620 & 0.9004 \\
8,21,21 & 21 & 85\% & $0.0447 \pm 0.0002$ & 9\% & $0.0287 \pm 0.0005$ & 0.0217 & 0.0690 & 0.0498 \\
& 21 & & $0.0441 \pm 0.0002$ & & $0.0285 \pm 0.0005$ & 0.0217 & 0.0690 & 0.0498 \\
\hline
& 5 & & $0.9598 \pm 0.0002$ & & $0.9749 \pm 0.0006$ & 0.9872 & 0.9386 & 0.9569 \\
5,15,25 & 15 & 68\% & $0.0309 \pm 0.0002$ & 6\% & $0.0197 \pm 0.0005$ & 0.0108 & 0.0464 & 0.0354 \\
& 25 & & $0.0093 \pm 0.0001$ & & $0.0054 \pm 0.0003$ & 0.0020 & 0.0150 & 0.0076 \\
\hline
\end{tabular}
	\caption{
    Escape probabilities for all mass sets. The columns \%(np) and \%(es) denote the percentage of systems satisfying the non-prompt ejections cut and the ergodic cut, respectively. $P_{s}^{(2d,np)}$ and $P_{s}^{(2d)}$ denote the escape probabilities (along with standard errors) measured from planar simulations after applying the non-prompt ejections cut and the ergodic cut, respectively. Kol 2D (Eq. \eqref{ep2}) and Kol 3D (Eq. \eqref{ep3}) denote the escape probability predictions from the Flux-based theory \citep{2021CeMDA.133...17K}. VK 2D denotes the escape probability predictions derived in \citep{2006tbp..book.....V}. For the rest of the theoretical predictions, refer to Table 2 of MKTL21.}
	\label{ept}
\end{table*}
%\end{landscape}

\begin{table*}[h]
	\centering
	\begin{tabular}{cccccc}
		\hline\hline
		Mass Set & Escaper & $N$ & $\alpha^{(sim)}$ & $L_{F,c}^{(sim)}$ & $L_{F,c}^{(th)}$  \\
		\hline
		& 15 (0)  & 62  & $1.234 \pm 0.284$ & 53.15  & 50.29 \\
		15,15,15 & 15 (1) & 56  & $1.249 \pm 0.300$ & 57.83  & 50.29 \\
		& 15 (2) & 59  & $1.070 \pm 0.270$ & 52.81 & 50.29 \\
		\hline
		& 12.5 & 168 & $1.358 \pm 0.182$ & 46.12  & 44.91  \\
		12.5,15,17.5 & 15 & 58 & $0.827 \pm 0.240$ & 171.31 & 170.25 \\
		& 17.5  & 38 & $0.803 \pm 0.292$ & 256.63  & 251.84 \\
		\hline
		& 12  & 208 & $1.107 \pm 0.146$ & 46.21  & 43.67  \\
		12,15,18 & 15 & 70 & $1.401 \pm 0.287$ & 201.39  & 197.55 \\
		& 18  & 50 & $1.261 \pm 0.320$ & 295.12  & 289.24 \\
		\hline
		& 10  & 435 & $0.832 \pm 0.088$ & 36.22  & 34.63  \\
		10,10,20 & 10 & 405 & $0.877 \pm 0.093$ & 36.27  & 34.63 \\
		& 20 & 63 & $0.722 \pm 0.217$ & 308.44 & 305.87 \\
		\hline
		& 10  & 337 & $0.910 \pm 0.104$ & 39.01  & 38.24  \\
		10,15,20 & 15  & 60 & $1.000 \pm 0.258$ & 321.34  & 316.05 \\
		& 20 & 57 & $1.002 \pm 0.265$ & 433.38  & 430.41  \\
		\hline
		& 10  & 289 & $1.097 \pm 0.123$ & 43.09  & 41.35 \\
		10,20,20 & 20 & 46 & $1.255 \pm 0.332$ & 548.19  & 542.70 \\
		& 20 & 56 & $0.950 \pm 0.261$ & 548.18 & 542.70 \\
		\hline
		& 8  & 359 & $0.990 \pm 0.105$ & 36.38  & 34.46 \\
		8,21,21 & 21 & 44 & $1.113 \pm 0.319$ & 698.79  & 695.27 \\
		& 21 & 48 & $0.726 \pm 0.249$ & 698.92 & 695.27 \\
		\hline
		& 5  & 510 & $1.005 \pm 0.089$ & 22.01  & 21.36 \\
		5,15,25 & 15 & 80 & $0.588 \pm 0.178$ & 653.71  & 652.67 \\
		& 25 & 52 & $1.038 \pm 0.283$ & 735.81  & 730.37 \\
		\hline
	\end{tabular}
    \caption{Estimated values and standard errors for the parameter $\alpha$ in \eqref{lf2de} using Maximum Likelihood Estimation for all the mass sets. The theoretical prediction is $\alpha=1$. $L_{F,c}^{(\mathrm{sim})}$ denotes the minimum value of $L_F$ measured from the simulations. $L_{F,c}^{(th)}$ denotes the theoretically predicted value of $L_{F,c}$. Column $N$ denotes the number of data points contained within the domain considered for estimation, $L_{F,c}\leq L_F \leq L_{F,c}+30$.}
    \label{mlet}
\end{table*}

\begin{table*}[h]
	\centering
	\begin{tabular}{cccccc}
		\hline\hline
		Mass set & Escaper & Pro/Ret Ratio & $P_{s}^{(pro)}$ & $P_{s}^{(ret)}$ & $P_{s}^{(2d)}$ \\
		\hline
		& 15 & & 0.326 & 0.341 &  0.332 \\
		15,15,15 & 15 & 0.90 & 0.336  & 0.331 & 0.334 \\
		& 15 & & 0.338 & 0.327 & 0.333 \\
		\hline
		& 12.5  &  & 0.628 & 0.487 & 0.560  \\
		12.5,15,17.5 & 15& 1.04 & 0.237 & 0.311  & 0.274  \\
		& 17.5  &  & 0.135 & 0.201 & 0.166 \\
		\hline
		& 12  &  & 0.676  & 0.519 & 0.603  \\
		12,15,18 & 15 & 1.14 & 0.218 & 0.298 & 0.256  \\
		& 18  &  & 0.106 & 0.183 &  0.141 \\
		\hline
		& 10  &  & 0.490  & 0.466 &  0.481 \\
		10,10,20 & 10 & 1.53 & 0.486 & 0.457 & 0.476  \\
		& 20 &  & 0.023 & 0.076 & 0.043 \\
		\hline
		& 10  &  & 0.844 & 0.654 & 0.783  \\
		10,15,20 & 15  & 2.08 & 0.114 & 0.222 & 0.149 \\
		& 20 & & 0.041 & 0.124 & 0.068 \\
		\hline
		& 10  &  & 0.917  & 0.706 & 0.874 \\
		10,20,20 & 20 & 3.85 & 0.042 & 0.145 &0.063  \\
		& 20 &  & 0.041 & 0.148 & 0.063 \\
		\hline
		& 8  &  & 0.960  & 0.734 & 0.943 \\
		8,21,21 & 21 & 12.35 & 0.020 & 0.136 & 0.029  \\
		& 21 &  & 0.020 & 0.130 & 0.028 \\
		\hline
		& 5  &  & 0.980  & 0.387 &0.975 \\
		5,15,25 & 15 & 144.00 & 0.016 & 0.493 &0.020  \\
		& 25 &  & 0.004 & 0.120 &0.005 \\
		\hline
	\end{tabular}
	\caption{
    Escape probabilities for prograde and retrograde escape types in the planar system. The column Pro/Ret Ratio denotes the ratio of systems undergoing prograde escapes to those undergoing retrograde escapes. Pro/Ret Ratio $=1$ is the theoretical prediction. $P_{s}^{(pro)}$ and $P_{s}^{(ret)}$ denote the escape probabilities for prograde and retrograde escapes, respectively. $P_{s}^{(2d)}$ denotes the overall escape probabilities measured from simulations.}
	\label{prp}
\end{table*}

\end{appendix}
\end{document}